# Electric field-induced creation and directional motion of domain walls and skyrmion bubbles

Chuang Ma,[1, †] Xichao Zhang,[2, †] Jing Xia,[2] Motohiko Ezawa,[3] Wanjun Jiang,[4]

Teruo Ono,[5] S. N. Piramanayagam,[6] Akimitsu Morisako,[1] Yan Zhou,[2, *] and Xiaoxi Liu [1, *]

[1]Department of Electrical and Computer Engineering, Shinshu University, 4-17-1 Wakasato, Nagano 380-8553, Japan

[2]School of Science and Engineering, The Chinese University of Hong Kong, Shenzhen, Guangdong 518172, China

[3]Department of Applied Physics, The University of Tokyo, 7-3-1 Hongo, Tokyo 113-8656, Japan

[4]State Key Laboratory of Low-Dimensional Quantum Physics and Department of Physics, Tsinghua University, Beijing 100084, China

[5]Institute for Chemical Research, Kyoto University, Gokasho, Uji, Kyoto 611-0011, Japan

[6]School of Physical and Mathematical Sciences, Nanyang Technological University, 637371, Singapore

†These authors contributed equally:
Dr. Chuang Ma and Dr. Xichao Zhang

*Corresponding Authors:
Dr. Yan Zhou
E-mail: zhouyan@cuhk.edu.cn
Dr. Xiaoxi Liu
E-mail: liu@cs.shinshu-u.ac.jp

(Dated: December 12, 2018)

## ABSTRACT

Magnetization dynamics driven by an electric field could provide long-term benefits to information technologies because of its ultralow power consumption. Meanwhile, the Dzyaloshinskii-Moriya interaction in interfacially asymmetric multilayers consisting of ferromagnetic and heavy-metal layers can stabilize topological spin textures, such as chiral domain walls, skyrmions, and skyrmion bubbles. These topological spin textures can be controlled by an electric field, and hold promise for building advanced spintronic devices. Here, we present an experimental and numerical study on the electric field-induced creation and directional motion of topological spin textures in magnetic multilayer films and racetracks with thickness gradient and interfacial Dzyaloshinskii-Moriya interaction at room temperature. We find that the electric field-induced directional motion of chiral domain wall is accompanied with the creation of skyrmion bubbles at certain conditions. We also demonstrate that the electric field variation can induce motion of skyrmion bubbles. Our findings may provide opportunities for developing skyrmion-based devices with ultralow power consumption.

The electric field (EF) induced spintronic phenomena could offer great benefits to the information-related industries, since they can be harnessed for building information processing devices with ultralow power consumption[1, 2]. Especially, the modification of magnetic parameters, such as the magnetic anisotropy[3-7], plays an essential role in realizing magnetization switching[5, 8-12], domain structure modification[13-15], and domain wall motion[16-21]. Additionally, recent experiments on magnetic asymmetric multilayers show that the antisymmetric Dzyaloshinskii-Moriya interaction (DMI), which can be induced at the ferromagnet/heavy metal interface[22-26], is able to further stabilize novel topological spin textures including chiral domain walls, skyrmions, and skyrmion bubbles. In a general context, the skyrmion bubble stands for the topologically non-trivial bubble with a fixed chirality but a larger size in compared with the compact skyrmion[24-26]. These topological spin textures also provide emerging opportunities for developing low-power information processing as well as high-density storage and logic computing technologies[22-29], which have been theoretically predicted and experimentally demonstrated in the last decade[22-25, 27].

The creation and motion of domain walls are indispensable processes in magnetic memory applications, where information are encoded by a sequence of movable domain walls representing binary bits. The mechanism is also applicable to information processing applications based on skyrmions and skyrmion bubbles[23-25, 30, 31]. Theoretical and experimental studies have suggested that topological spin textures can be created and driven in multiple different ways, such as by applying magnetic fields or electric currents[22-26, 32-37]. However, methods based on an electric current may be energy consuming and produce significant Joule heating, which impede the integration into actual circuits with nanoscale dimensions. Therefore, as the Joule heating effect can be significantly suppressed under an EF, the use of an EF appears to be an efficient and robust method for creating, driving, and controlling magnetic structures[2, 38-47].

The EF-controlled or EF-assisted motion of a domain wall has already been realized[16-20], which shows that the domain wall velocity can be changed by the EF (applied voltage $V = \pm 2 \sim \pm 10$ V) effectively. The influence of a sloped EF on the domain wall creep motion has been studied[48], which suggests the sloped EF may be used to assist the domain wall motion. The domain structure can also be significantly modified by the EF ($V = \pm 10$ V)[15]. Most recently, it was found that a local EF ($V = \pm 3$ V) can be used to create a skyrmion from the ferromagnetic background at low temperature[38]. The creation and annihilation of skyrmion bubbles at room temperature by the EF ($V = \pm 20$ V) have also been demonstrated[39]. In addition, recent reports show it is possible to control the skyrmion chirality[40] and control the transition between the helical phase and skyrmion phase using an EF[49].

In this paper, we experimentally demonstrate the EF-induced creation and directional motion of a chiral domain wall and skyrmion bubbles in a magnetic multilayer film as well as a multilayer racetrack with thickness gradient and interfacial DMI at room temperature, where the EF is directly applied at the metal (Pt)/dielectric ($SiO_2$) interface. Note that the metal/dielectric interface has recently been suggested as a unique system to provide a large EF-induced magnetic anisotropy change due to the electric quadrupole induction[7], which differs from the case where the magnetic anisotropy is modified by the EF-induced charge doping at the ferromagnet/dielectric interface[4-6]. We found that the motion of the chiral domain wall is accompanied with the creation of skyrmion bubbles in the presence of certain out-of-plane magnetic field, which acts as a new method for creating skyrmion bubbles. Micromagnetic simulations help to understand the experimental observations, which indicate that the spatially varied anisotropy profile generated by the thickness gradient, the EF-induced anisotropy change, as well as the DMI play essential roles in the observed domain wall dynamics. Our results are useful for the development of next-generation EF-controlled low-power information devices.

Figure 1a shows the experimental setup, where a [Pt (0.5 nm)/CoNi (0.5 nm)/Pt (0.5 nm)/CoNi (0.5 nm)/Pt (1.0 nm)] multilayer is sandwiched between the indium tin oxide (ITO)/$SiO_2$ bilayer and the glass substrate (see Methods). Due to the broken inversion symmetry at the ferromagnet/heavy metal interfaces[23-25, 50], certain interfacial DMIs are expected to be generated at the Pt/CoNi interfaces. Indeed, based on the asymmetric domain wall motion experiment, the magnitude of DMI is measured to be $D = 0.08 \sim 0.12$ mJ/$m^2$ in our samples (see Fig. S1 and Fig. S2 in the Supporting Information). The multilayer structure is fabricated to have a 20-$\mu$m-long thickness gradient from the left edge along the $x$ direction, which means that the total thickness of the multilayer decreases from 3 nm at $x \sim 20$ $\mu$m to zero at $x \sim 0$ $\mu$m (see Fig. S3 in the Supporting Information). As shown in Fig. 1b, in addition to the multilayer film, we also fabricate some racetracks with thickness gradient at left ends based on the multilayer structure given in Fig. 1a. Figure 1c schematically illustrates the EF-controlled magnetic spin texture transition in the multilayer film and racetrack, which will be discussed below.

We first study the properties and EF-induced domain wall dynamics of the multilayer film. Figure 2a shows the experimentally measured out-of-plane hysteresis loops at different locations with different thicknesses using the magneto-optical Kerr effect (MOKE) technique at room temperature (see Methods). The horizontal distance between the left edge and the Kerr laser spot is defined as $l$. Note that the diameter of the laser spot is determined to be 3 $\mu$m. The decreasing thickness toward the left edge results in a decreasing remanent magnetization $M_R$. It can be seen that the value of $M_R/M_S$ increases with $l$ and reaches a constant value when $l > 25$ $\mu$m (see Fig. 2a inset), which indicates that the thickness gradient leads to a transition region near the left edge. The coercivity field decreases as the thickness of sample decreases (see Fig. S4 in the Supporting Information), which indicates that the perpendicular magnetic anisotropy (PMA) is reduced for the region with thickness gradient while the region without the thickness gradient has almost constant PMA. The presence of the DMI naturally results in the tilted edge magnetization[23-25], which further leads to the formation of an in-plane domain near the left edge ($x \sim 0$ $\mu$m) due to the reduced PMA in the vicinity of the left edge (see Fig. 1c and Fig. S5 in the Supporting Information).

Figure 2b shows the room-temperature out-of-plane hysteresis loops measured by the MOKE technique at $l = 30$ $\mu$m for different amplitudes of the EF. It shows a slight reduction in coercivity when a gate voltage $V_G$ is applied, either positive (+4 V) or negative (-4 V). Here, the negative $V_G$ is defined as an electron accumulation at the top surface of the multilayer. When a larger $V_G$ is applied (-10 V), the measured Kerr signal shows a significantly reduced $M_R$ as well as a deformation of the hysteresis loop, which indicates a decrease of the PMA in the sample. The inset of Fig. 2b shows $M_R/M_S$ as a function of $V_G$. We also show out-of-plane hysteresis loops and $M_R/M_S$ as a function of $V_G$ for different $l$ in Fig. S6 in the Supporting Information and the coercivity field as a function of $V_G$ for $l = 40$ $\mu$m in Fig. S7 in the Supporting Information. It can be seen that when $V_G < -6$ V, the PMA of the sample significantly reduces, implying that the negative $V_G$ induced electron accumulations at the Pt/SiO$_2$ interface could result in the reduced PMA. It is worth mentioning that the effects of the positive and negative voltage effects are symmetric (see Fig. S7 and Fig. S8 in the Supporting Information). In this work, we only focus on the case of low negative $V_G$ ($|V_G| < 10$ V), which is comparable to or smaller than the $V_G$ used in recent studies (cf. Refs. [16-20, 38] and 33).

Figure 3 shows the top-view MOKE images of the multilayer film region with the thickness gradient ($x < 20$ $\mu$m), where the initial out-of-plane magnetization is pointing along the $+z$ direction. In Fig. 3a, we first apply the $V_G$ in the presence of an external magnetic field of $H_z = -0.2$ mT (see Movie 1 in the Supporting Information). It shows that for $V_G = -2$ V no domain wall is created and the sample remains in the initial magnetization configuration. For $V_G = -6.5$ V two chiral domain walls are created in the vicinity of the left edge, and the right chiral domain wall moves toward the $+x$ direction, resulting in the formation of a spin-down (pointing at the $-z$ direction) domain. When $V_G$ is increased to -9 V, the right chiral domain wall moves further rightwards and the displacement of the chiral domain wall reaches about 10 $\mu$m. Indeed, the spin-down domain expands when the right chiral domain wall moves rightwards. It is worth mentioning that during the EF-induced motion of the chiral domain wall, we observe the creation of many circular domain wall structures near the right domain wall (see the red box area in Fig. 3a), of which the diameters are about 0.5 ~ 1 $\mu$m. Based on the observation as well as the presence of DMI at the Pt/CoNi interfaces[50] (see Fig. S1 and Fig. S2 in the Supporting Information), these circular domain wall structures should well be topologically non-trivial skyrmion bubbles with fixed chirality, which are created due to the magnetization fluctuation caused by the electric field-induced anisotropy change and stabilized by a delicate competition between DMI, PMA and applied field (see Fig. S9 in the Supporting Information). We show numerically in Fig. S10 in the Supporting Information that the presence of certain small DMI (e.g. 0.05 mJ m$^{-2}$) is enough to induce and fix the chirality of a micrometer-sized bubble, leading to the formation of skyrmion bubble with an integer topological charge.

Moreover, our asymmetric domain wall expansion experiment manifests the Néel-type chiral nature of domain wall structures in our sample (see Fig. S1 and Fig. S2 in the Supporting Information).

These skyrmion bubbles also move with the right chiral domain wall when $V_G$ is varied. When $V_G$ is reduced back to -6.5 V from -9 V, the right chiral domain wall moves toward the -x direction. Meanwhile, many skyrmion bubbles remain as stable structures in the vicinity of the right chiral domain wall, which indicates the stability and rigidity of these observed skyrmion bubbles. When $V_G$ is further reduced to -4.5 V, the right chiral domain wall moves very close to the left chiral domain wall, and the spin-down domain shrinks and almost disappears. However, skyrmion bubbles still exist due to their topologically protected stability. It should be noted that all skyrmion bubbles will be annihilated when $V_G$ is reduced to zero or switched off. Figure 3b and 3c show the EF-induced creation and motion of the chiral domain wall at $H_z = 0$ mT and $H_z = +0.2$ mT (see Movie 2 and Movie 3 in the Supporting Information). It shows an almost identical EF-induced domain wall motion compared to Fig. 3a, yet without the accompanying creation of skyrmion bubbles. The reason could be that a certain magnetic field applied pointing along the -z-direction is helpful for increasing the size and thus the stability of the skyrmion bubble, of which the core magnetization is aligned along the -z-direction in the given sample. For comparison purpose, we also study the magnetic field-induced motion of the chiral domain wall in Fig. 3d by applying an increasing magnetic field along the -z direction (see Movie 4 in the Supporting Information), where the magnetization is pointing along the +z direction initially. It is found that the magnetic field-induced motion of the chiral domain wall is in stark contrast to that induced by the EF. First, when the magnetic field is increasing in the -z direction, many domains with inner core magnetization pointing along the -z direction are formed near the left edge, which then expand in an asymmetric manner toward the +x direction. The magnetization of the sample is fully reversed when the magnetic field is increased to $H_z = -2.89$ mT. No skyrmion bubbles are found during the magnetic field-induced motion of domain walls.

In order to demonstrate the repeatability of the EF-induced chiral domain wall motion, we adjust the Kerr laser spot to focus on a certain location ($l = 30$ $\mu$m), where the chiral domain wall is able to reach the Kerr laser spot area, and then apply an alternating $V_G$, as depicted in Fig. 4a. The maximum and minimum amplitudes of $V_G$ are set as -8 V and 0 V, respectively. The frequency of the applied alternating $V_G$ is set as 7 Hz. At a large $V_G$, the right chiral domain wall moves forwards, resulting in a spin-down domain underneath the Kerr laser spot. Similarly, at a small $V_G$, the right chiral domain wall moves backwards, resulting in a spin-up domain underneath the Kerr laser spot. Therefore, the observed Kerr signal is oscillating, as a response to the repetitive motion of the chiral domain wall driven by the alternating $V_G$, as shown in Fig. 4b. We also apply the electric field with different rate of change and the results are given in Fig. S11 in the Supporting Information. This demonstrates the repeatability of the EF-induced chiral domain wall motion as well as the possibility of using the EF-induced chiral domain wall motion in building binary memory devices. It should be noted that it is difficult to observe the domain wall dynamics at the immediate vicinity of the left edge ($l \sim 0$ $\mu$m) as the magnitude of the MOKE signal decreases rapidly with decreasing $l$. For the purpose of creating more than one chiral domain wall in the system, one could fabricate multiple racetracks in one system or fabricate a single racetrack with multiple regions with periodic thickness gradients (see Fig. S12 in the Supporting Information).

For the purpose of device applications, we continue to demonstrate the EF-induced motion of chiral domain wall and skyrmion bubbles in a racetrack with film thickness gradient at the left end. As shown in Fig. 5, the width of the racetrack equals 6 $\mu$m and the total length of the racetrack is longer than 500 $\mu$m. The external applied magnetic field is equal to zero, i.e. $H_z = 0$ mT. When the EF is applied to the racetrack with $V_G$ increasing from 0 V to -4 V in a step-by-step manner (see Fig. 5a), a reversed domain is formed near the left end of the racetrack. The created domain wall propagates toward the right as $V_G$ further increases from -4 V to -7 V. The domain wall displacement change reaches ~20 $\mu$m at $V_G$ = -7 V, of which is determined as the spacing between the domain wall positions at $V_G$ = -6 V and $V_G$ = -7 V. During the motion of the domain wall, several skyrmion bubbles are created in the front of the domain wall. When $V_G$ is reduced from -7 V to -3.5 V, both the domain wall and skyrmion bubbles move toward the left end of the racetrack, where the speed of the skyrmion bubbles is much slower than that of the domain wall (see Movie 5 in the Supporting Information). We can see that the skyrmion Hall effect is not obvious as our racetrack sample can be treated as a confined system due to the narrow track width and large skyrmion bubble size, where the geometric confinement effect (i.e., the skyrmion-edge repulsion) results in the skyrmion motion along the longitudinal direction (see Figs. S13-S15 and section S1 in the Supporting Information). It should be noted that some skyrmion bubbles are pinned or annihilated during the motion toward the left end when $V_G$ is decreasing. The pinning or annihilation of skyrmion bubbles are usually resulted by defects or impurities (see Fig. S16 and Fig. S17 in the Supporting Information). Figure 5 also shows the domain wall displacement change and average speed in the racetrack as functions of $V_G$ and time. It can be seen that the domain wall displacement change is proportional to the absolute value of $V_G$, where the maximum speed of the domain wall motion induced by the voltage change could reach 50 mm s$^{-1}$ (see Fig. 5f). It should be mentioned that the domain wall and skyrmion bubbles can move forwards and backwards in our experiments. For the domain wall, the forward and backward motion are mainly induced by the change of PMA gradient. However, for the skyrmion bubble, the mechanisms for the forward and backward motion are different. Namely, the forward motion of skyrmion bubbles is induced by the push of a forward moving domain wall due to the repulsion between the domain wall and skyrmion bubbles. In contrast, the backward motion of skyrmion bubbles is directly driven by the PMA gradient as the PMA gradient can provide a driving force on skyrmion bubbles (see Fig. S18 and section S2 in the Supporting Information).

Having experimentally demonstrated the EF-induced motion of the chiral domain wall and skyrmion bubble, we further study these observed phenomena in a qualitative manner under the framework of micromagnetics (see Methods). Figure 6 shows the numerical simulation results on the EF-induced creation and motion of the chiral domain wall. The geometry of the simulated sample is $600 \times 300 \times 0.5$ nm$^3$. For the purpose of modeling the effect of the thickness gradient, we also assume a linearly increasing PMA along the $x$ direction from $x = 0$ nm to $x = 450$ nm. As shown in Fig. 6a, we assume that the PMA linearly increases from $K_u = 0$ at $x = 0$ nm to $K_u = K$ at $x = 450$ nm in the absence of the EF. When the $V_G$ is applied, we assume that the PMA linearly increases from $K_u = 0$ at $x = 0$ nm to $K_u = 0.6K$ at $x = 450$ nm, showing the reduced PMA induced by the EF but regardless of the exact value of $V_G$. When the amplitude of $V_G$ is further increased, we assume that the PMA linearly increases from $K_u = 0$ at $x = 0$ nm to $K_u = 0.55K$ at $x = 450$ nm. The detailed values of $K$ and other parameters used in the simulations are given in the Methods. It should be noted that we assumed a fixed small value of DMI in our simulations for simplicity as we expect the motion of chiral domain wall and skyrmion bubble are dominated by the change of the PMA. Recent theoretical studies have suggested that the magnitude of DMI can be modulated using an electric field[51]. Indeed, a recent experimental report has demonstrated that the skyrmion chirality can be controlled by voltage tuning of DMI[40].

In Fig. 6b, we simulated the EF-induced creation and motion of a chiral domain wall at $H_z = 0$ mT. At the initial state, the relaxed magnetization is pointing along the $+z$ direction, except the magnetization near the left edge, where it lies in the $xy$-plane and is pointing along the $+x$ direction due to the DMI and reduced PMA (see Fig. S5 in the Supporting Information). When the PMA is decreased due to the application of the EF (change #1), a chiral domain wall is formed and moves toward the $+x$ direction. When the PMA is further decreased (change #2), the domain wall continues to move toward the $+x$ direction. When the PMA is increased back to its original value (changes #3 and #4), the chiral domain wall moves toward the $-x$ direction and annihilates. It can be seen that the above simulated results agree well with the experimental observed chiral domain wall creation and motion induced by the EF (see Fig. 3). However, during the chiral domain wall motion process, no chiral bubbles are found, which differs from the experimental results. The reason could be that both the sizes of the experimental sample and the experimentally created skyrmion bubbles are much larger than those in the simulation, so that a large dipolar interaction in the experimental sample can help to excite and stabilize chiral bubbles. Besides, the defects and impurities in the real experimental sample could serve as nucleation points to promote the creation of skyrmion bubbles[36]. Hence, we slightly adjust the values of PMA and exchange constant so that a skyrmion bubble with the topological charge of $Q = -1$ could be created (see Methods). As shown in Fig. 6c, when we decrease (changes #1 and #2) and then increase the PMA (changes #3 and #4), a skyrmion bubble is created and moves forwards and backwards accordingly, which proves that the EF-induced PMA change can result in the motion of the skyrmion bubble, as both the chiral domain wall and skyrmion bubble are stabilized at a certain anisotropy value. The location corresponding to the certain anisotropy value shifts when the PMA profile is changed by the EF. We note that our simulation model does not take into account the change of the electric field with sample thickness for the sake of simplicity. Namely, we assumed that the wedge structure provides the PMA gradient while the electric field lowers the PMA in a uniform manner. In real samples, the electric field can be non-uniform along the wedge structure due to the short screening length of metals, so that its effect on the PMA could be reduced in the thicker part of the wedge relative to the thinner part.

In conclusion, using the MOKE technique, we have realized and observed the EF-induced creation and directional motion of a chiral domain wall and skyrmion bubbles in a magnetic multilayer film and a multilayer racetrack with interface-induced DMI and artificial thickness gradient at room temperature. It is found that a domain wall can be created and directionally displaced about 10 $\mu$m for a voltage as low as 9 V in the magnetic film. The EF-induced chiral domain wall motion is also demonstrated to be repeatable. Besides, we find that the EF-induced motion of the chiral domain wall is accompanied with the creation of skyrmion bubbles at a certain external out-of-plane magnetic field. In addition, we experimentally demonstrate that the domain wall displacement and velocity induced by the voltage change are proportional to the absolve value of the applied voltage. Micromagnetic simulations based on the EF-induced anisotropy change are used to help us to understand the experimental observations, implying the importance of the thickness gradient-induced anisotropy profile, the EF-induced anisotropy change, and the DMI on the domain wall and skyrmion bubble dynamics. Our findings suggest that magnetic multilayer films and racetracks with an artificial gradient in thickness, PMA, and DMI can be harnessed for designing EF-controlled magnetic devices with ultralow energy consumption.

**Methods.** *Experimental details.* The Pt (0.5 nm)/CoNi (0.5 nm)/Pt (0.5 nm)/CoNi (0.5 nm)/Pt (1.0 nm) films were deposited by DC magnetron sputtering under an Ar pressure of 0.35 Pa after the chamber was evacuated to a base pressure of about $5 \times 10^{-5}$ Pa. Racetrack trenches (6-$\mu$m-wide and 300-$\mu$m-long) were fabricated by a maskless photolithography system onto glass substrate with a 500-nm-thick photoresist (AZ 5200). Developed substrates were covered by a Ni stencil mask with a window of 500-$\mu$m-wide and 500-$\mu$m-long. The mask was carefully aligned and attached by a home designed magnet to cover one end of the racetrack trenches. A thickness gradient at the edges of the magnetic stripe was formed during sputtering due to the thickness of the stencil mask. A magnetic microwire with one end of wedge was first fabricated by photolithography and stencil mask technique as illustrated in left panel of Fig. S3 in the Supporting Information. A photoresist trench with depth of 0.5 μm was first developed by normal photolithography technique. A Ni stencil mask with thickness of 20 μm was carefully placed to cover one end of the photoresist trench. In a magnetron sputtering system, the sputtered atoms with initial velocity varies in a wide dispersion angles. Due the thickness of the stencil mask, a large amount of sputtered atoms will be shadowed by the edge of the stencil mask. A wedge structure will be formed after remove the stencil mask and lift-off the photoresist. It should be noted that we can control the profile of the wedge by simply varying the thickness of the stencil mask or by varying the target to substrate distance in the sputtering system. In this experiment, we are using stencil mask with thickness of 20 μm and a target to substrate distance of 70 μm. On the other hand, one side of the magnetic stripe was connected with Ta (3 nm)/Cu (100 nm)/Ta (10 nm). After lift-off the photoresist from the substrate, a 100-nm-thick $SiO_2$ film was deposited by RF sputtering under Ar + 10% $O_2$ pressure of 0.2 Pa to cover the magnetic stripe. ITO pad, with a thickness of 5 nm, were deposited above the $SiO_2$ and magnetic stripe by DC magnetron sputtering under Ar + 10% $O_2$ pressure of 0.4 Pa.

*Domain observation details.* Domain structures were observed by a Kerr microscope. A mercury lamp of 100 W is used to get sufficient contrast. The Kerr microscope use a 50X object lens with a typical resolution of 0.3 μm. In this experiment, the Kerr microscope are configured to observe polar Kerr effect.

***Local area out-of-plane hysteresis measurement.*** To measure the local area out-of-plane hysteresis loops at different part of the wedges, we have set-up a micro Laser polar Kerr effect measurement system. In this system, a circular polarized laser with wavelength of 532 nm were focused onto a spot of around 0.4 μm by a high resolution object lens of 50X. Such a spot size can secure local area out of plane hysteresis measurement at different part of a wedge which the thickness of changes from almost zero to 3 nm (total thickness is 3 nm with total CoNi thickness of only 1.0 nm) in a horizontal distance of 20 μm. The polarization of the reflected Laser beam was detected by a differential beam detector with magnification of 60 dB.

***Simulation details.*** The micromagnetic simulation is carried out by using the 1.2a5 release of the Object Oriented MicroMagnetic Framework (OOMMF)[52]. The simulator used a number of the standard OOMMF extensible solver (OXS) objects for modeling different micromagnetic energy terms. The OXS extension module for the interface-induced DMI is also used in our simulation. The three-dimensional (3D) time-dependent magnetization dynamics are described by the Landau-Lifshitz-Gilbert (LLG) equation, given as

$$\frac{d\boldsymbol{M}}{dt} = -\gamma_0 \boldsymbol{M} \times \boldsymbol{H}_{\text{eff}} + \frac{\alpha}{M_S}\left(\boldsymbol{M} \times \frac{d\boldsymbol{M}}{dt}\right), \tag{1}$$

where $\boldsymbol{M}$ is the magnetization, $M_S = |\boldsymbol{M}|$ is the saturation magnetization, $t$ is the time, $\gamma_0$ is the absolute gyromagnetic ratio, and $\alpha$ is the Gilbert damping constant. $\boldsymbol{H}_{\text{eff}} = -\delta\varepsilon/\mu_0\delta\boldsymbol{M}$ is the effective field, where the average energy density $\varepsilon$ contains the Heisenberg exchange, DMI, PMA, applied magnetic field, and demagnetization energy terms, which reads

$$\varepsilon = A\left[\nabla\left(\frac{\boldsymbol{M}}{M_S}\right)\right]^2 - K_u\frac{(\boldsymbol{n}\cdot\boldsymbol{M})^2}{M_S^2} - \mu_0\boldsymbol{M}\cdot\boldsymbol{H} - \frac{\mu_0}{2}\boldsymbol{M}\cdot\boldsymbol{H}_{\text{d}}$$

$$+ \frac{D}{M_S^2}[M_z(\boldsymbol{M}\cdot\nabla) - (\nabla\cdot\boldsymbol{M})M_z], \tag{2}$$

where $A$, $K_u$, and $D$ denote the Heisenberg exchange, PMA, and DMI constants, respectively. $\boldsymbol{H}$ is the applied external magnetic field, and $\boldsymbol{H}_{\text{d}}$ is the demagnetization field. $M_z$ is the out-of-plane Cartesian component of the magnetization $\boldsymbol{M}$. $\mu_0$ is the vacuum permeability constant, and $\boldsymbol{n}$ is the unit surface normal vector. In the simulation, we treat the model as a single layer, where the lateral cell size is set as $5 \times 5$ nm$^2$. The saturation magnetization is measured to be $M_S = 400$ kA m$^{-1}$ for the region without the thickness gradient. For simulation of the domain wall, we used parameters: $\alpha = 0.2$, $A = 10$ pJ m$^{-1}$, $D = 0.5$ mJ m$^{-2}$, and $K = 0.214$ MJ m$^{-3}$. For simulation of the skyrmion bubble, we used parameters: $A = 8$ pJ m$^{-1}$ and $K = 0.218$ MJ m$^{-3}$. In our simulations, the effect of the EF is assumed to induce a change of the PMA. The topological magnetization textures are characterized by the topological charge given as

$$Q = \frac{1}{4\pi} \int \boldsymbol{m} \cdot \left( \frac{\partial \boldsymbol{m}}{\partial x} \times \frac{\partial \boldsymbol{m}}{\partial y} \right) dx dy, \tag{3}$$

where $\boldsymbol{m} = \boldsymbol{M}/M_S$ is the reduced magnetization. The topological charge $Q$ is also referred to as the skyrmion number. A topologically trivial bubble has $Q = 0$, while a topologically non-trivial ground-state skyrmion has $Q = \pm 1$.

## ASSOCIATED CONTENT

### Supporting Information

The Supporting Information is available free of charge on the ACS Publications website at DOI:

Theoretical description of the skyrmion Hall effect, estimation of the driving force generated by a PMA gradient, asymmetric circular domain wall expansion, illustration for the fabrication of thickness gradient, dependence of coercivity on film thickness, simulated relaxed states at different DMI, hysteresis loops and remanences measured at different positions, coercivity fields measured at different voltage, simulated skyrmion bubble creation, simulated relaxation of a bubble in the absence and presence of DMI, dependence of Polar Kerr signal on voltage sweep rate, simulated EF-induced creation and motion of domain walls, Kerr microscope images at the wedge structure, simulated skyrmion motion in narrow and wide racetracks, simulated defect-induced pinning and annihilation of a skyrmion, illustration for the EF-induced skyrmion motion, compact skyrmion and skyrmion bubble profiles. (PDF)

Movie 1: Electric field-induced motion of a chiral domain wall accompanied with creation of chiral skyrmion bubbles at $H_z$ = -0.2 mT. (AVI)

Movie 2: Electric field-induced motion of a chiral domain wall at $H_z$ = 0 mT. (AVI)

Movie 3: Electric field-induced motion of a chiral domain wall at $H_z$ = +0.2 mT. (AVI)

Movie 4: Magnetic field-induced motion of a chiral domain wall in the absence of the electric field. (AVI)

Movie 5: Electric field-induced motion of a chiral domain wall accompanied with creation of chiral skyrmion bubbles in a racetrack. (AVI)

## AUTHOR INFORMATION

### Corresponding Author

*E-mail: zhouyan@cuhk.edu.cn

*E-mail: liu@cs.shinshu-u.ac.jp

**Author Contributions**

Chuang Ma and Xichao Zhang contributed equally to this work.

**Notes**

The authors declare no competing financial interest.

**ACKNOWLEDGEMENTS**

X.Z. acknowledges the support by the Presidential Postdoctoral Fellowship of the Chinese University of Hong Kong, Shenzhen (CUHKSZ). M.E. acknowledges the support by the Grants-in-Aid for Scientific Research from JSPS KAKENHI (Grant Nos. JP18H03676, JP17K05490 and JP15H05854), and also the support by CREST, JST (Grant Nos. JPMJCR16F1 and JPMJCR1874). W.J. was supported by the National Key R&D Program of China (Grant Nos. 2017YFA0206200 and 2016YFA0302300), the National Natural Science Foundation of China (Grant Nos. 11774194 and 51831005), the 1000-Youth talent program of China, the State Key Laboratory of Low-Dimensional Quantum Physics, and the Beijing Advanced Innovation Center for Future Chip (ICFC). T.O. acknowledges the support by the Grants-in-Aid for Scientific Research from JSPS KAKENHI (Grant No. 15H05702). S.N.P. acknowledges the support by NTU-JSPS research grant. Y.Z. acknowledges the support by the President's Fund of CUHKSZ, the National Natural Science Foundation of China (Grant No. 11574137), and Shenzhen Fundamental Research Fund (Grant Nos. JCYJ20160331164412545 and JCYJ20170410171958839). X.L. acknowledges the support by the Grants-in-Aid for Scientific Research from JSPS KAKENHI (Grant Nos. 17K19074, 26600041 and 22360122).

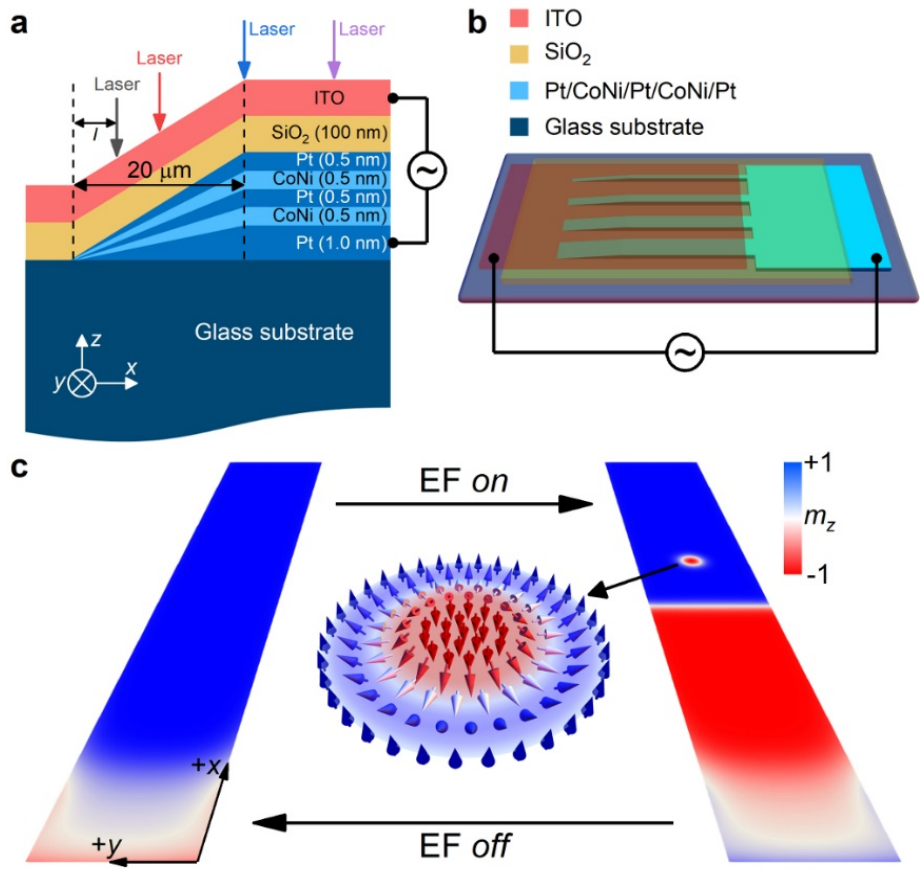

**Figure 1. Illustrations of the sample configuration and magnetic spin textures transition controlled by the EF. a**, Illustration of the experimental film, which has a 20-$\mu$m-long artificial thickness gradient along the $x$ direction. The length between the MOKE laser spot (used for measuring the hysteresis loop) and the left edge, where the thickness equals zero, is defined as $l$. The $V_G$ is applied between the multilayer and the ITO as indicated. **b**, 3D illustration of the experimental samples, which have artificial thickness gradients at the left ends. **c**, Illustration of the EF-controlled magnetic spin textures transition in the film. When the EF is applied, a domain wall will be created and propagate in the magnetic film. At certain conditions, chiral bubbles will also be created during the application of the EF.

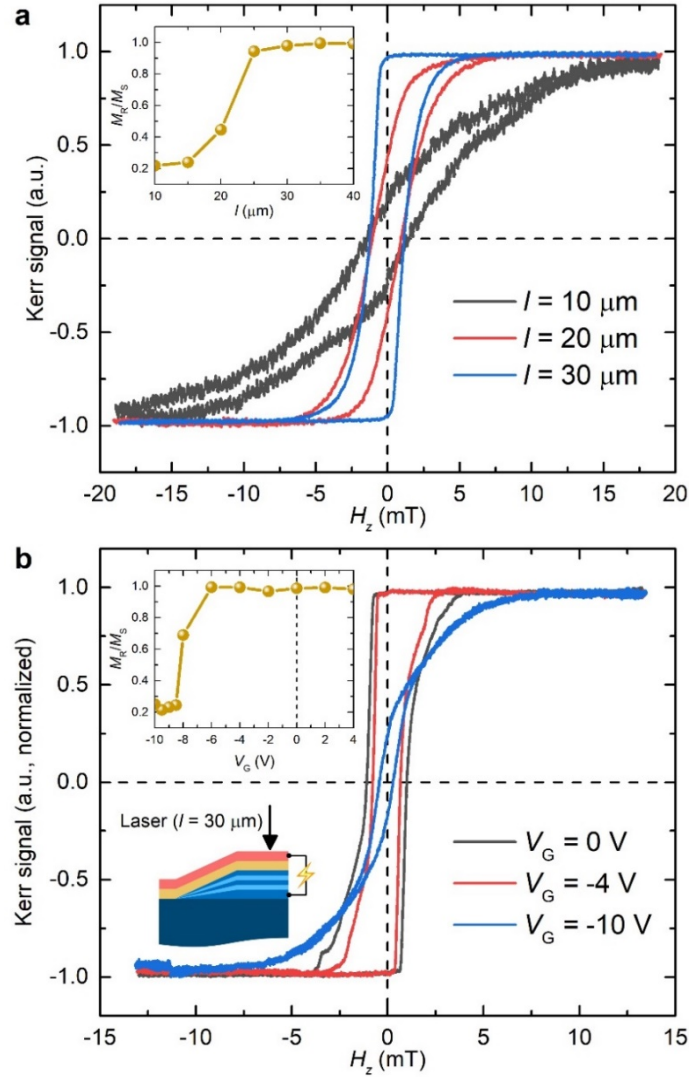

**Figure 2. Hysteresis loops at different thicknesses for $V_G$ = 0 V and hysteresis loops for different $V_G$. a**, Hysteresis loops measured by the MOKE technique at different locations along the $x$ direction at $V_G$ = 0 V. Inset shows $M_R/M_S$ as a function of $l$, which indicates the reduced PMA near the edge at $x \sim 0$ $\mu$m and the PMA at $x > 25$ $\mu$m. **b**, Hysteresis loops measured by the MOKE technique at $l$ = 30 $\mu$m for different $V_G$. The Kerr signal has been normalized. The negative $V_G$ result in a decrease of the coercivity. Inset shows $M_R/M_S$ as a function of $V_G$, which indicates that the PMA can be significantly reduced for $V_G$ < -6 V.

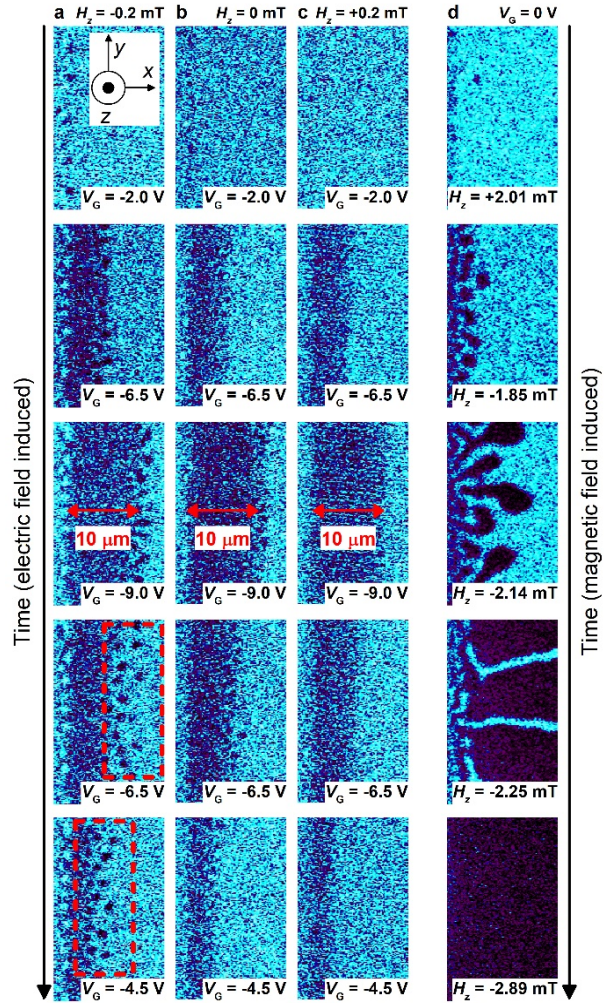

**Figure 3. MOKE microscopy images of the EF-induced and magnetic-field-induced motion of chiral domain walls and skyrmion bubbles in the film with the thickness gradient. a**, The EF-induced creation and motion of a chiral domain wall accompanied by the creation of skyrmion bubbles at $H_z$ = -0.2 mT. The red box indicates the area where many skyrmion bubbles are created. **b**, The EF-induced creation and motion of a chiral domain wall and skyrmion bubbles at $H_z$ = 0 mT. The size of the skyrmion bubbles is smaller than that in **a**. **c**, The EF-induced creation and motion of a chiral domain wall at $H_z$ = +0.2 mT. **d**, The magnetic-field-induced chiral domain wall motion at $V_G$ = 0 V. Note that the tiny dark and bright spots in the saturated state are caused by noise. The window focuses on the area with thickness gradient.

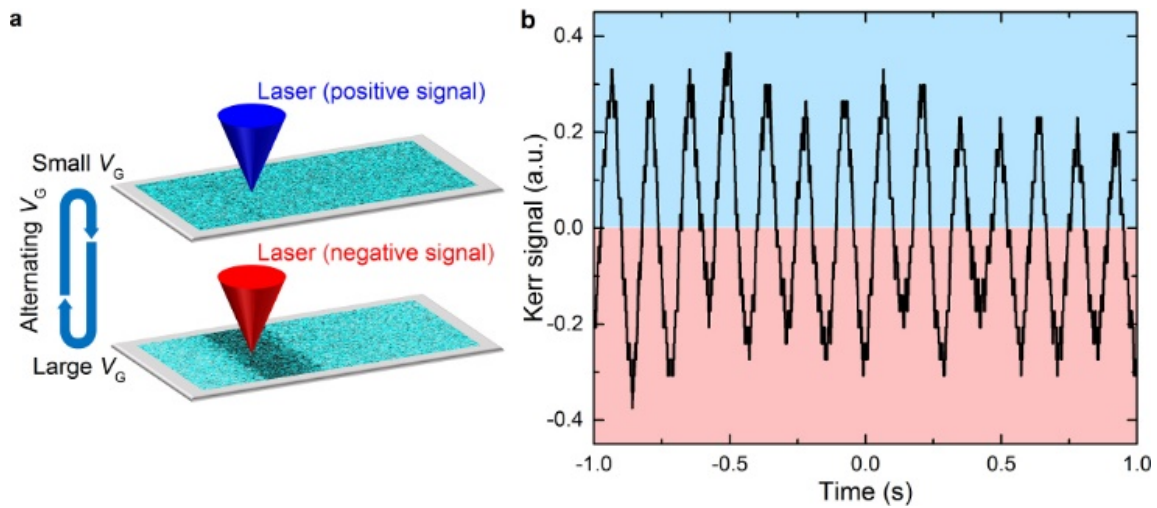

**Figure 4. Repetitive motion of the domain wall driven by an alternating EF. a**, Schematic drawing showing the location of the MOKE laser spot as well as the displacement of the chiral domain wall induced by the EF. A large $V_G$ results in a spin-down region underneath the laser spot, while a small $V_G$ results in a spin-up region underneath the laser spot. **b**, Time-dependent Kerr signals corresponding to the input alternating $V_G$, which indicate that the chiral domain wall moves continuously backwards and forwards.

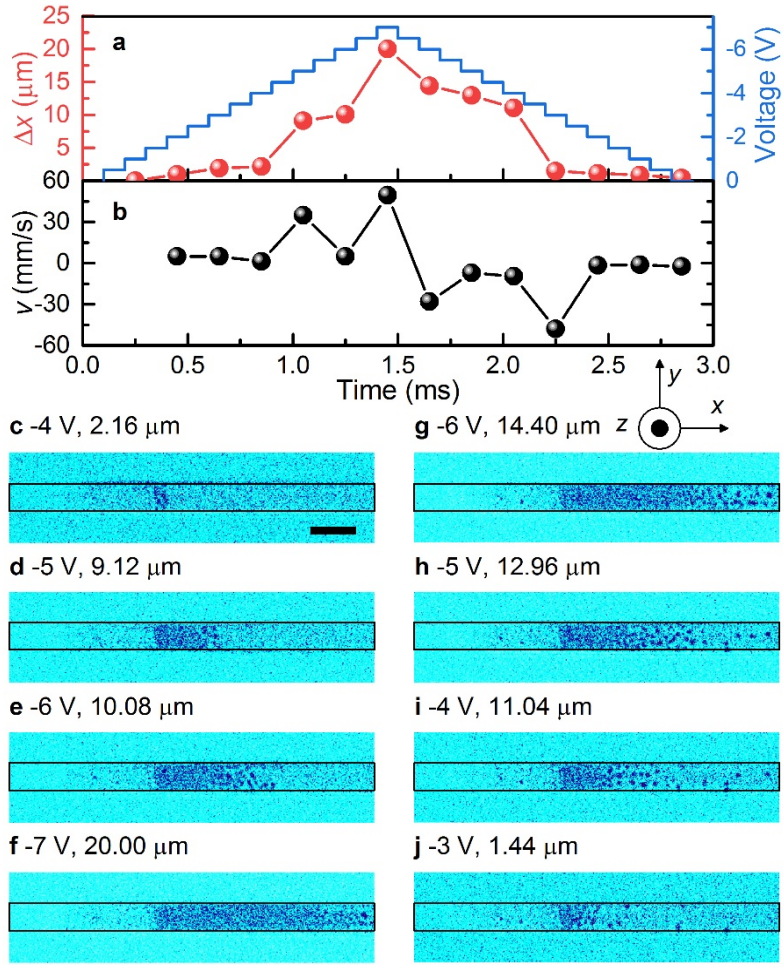

**Figure 5. EF-induced motion of chiral domain walls and skyrmion bubbles in the racetrack with the thickness gradient. a**, Domain wall displacement change and applied voltage ($V_G$) as functions of time. **b**, the average velocity of the domain wall for each change of the applied voltage. The EF is applied to the 6-$\mu$m-wide racetrack at $H_z = 0$ mT, where $V_G$ increases from 0 V to -7 V and then reduced to 0 V in a step-by-step manner with a step variation of 0.5 V per 0.1 ms. The domain wall displacement change is determined as the spacing between the domain wall positions before and after one-volt variation of $V_G$. When $V_G < 2$ V, no domain wall motion is observed. **c-j**, MOKE microscopy images of the racetrack at selected times. The values of $V_G$ and domain wall displacement change are indicated above each MOKE image. The black box indicates the racetrack shape. Note that the MOKE images only show part of the racetrack. Scale bar, 10 $\mu$m.

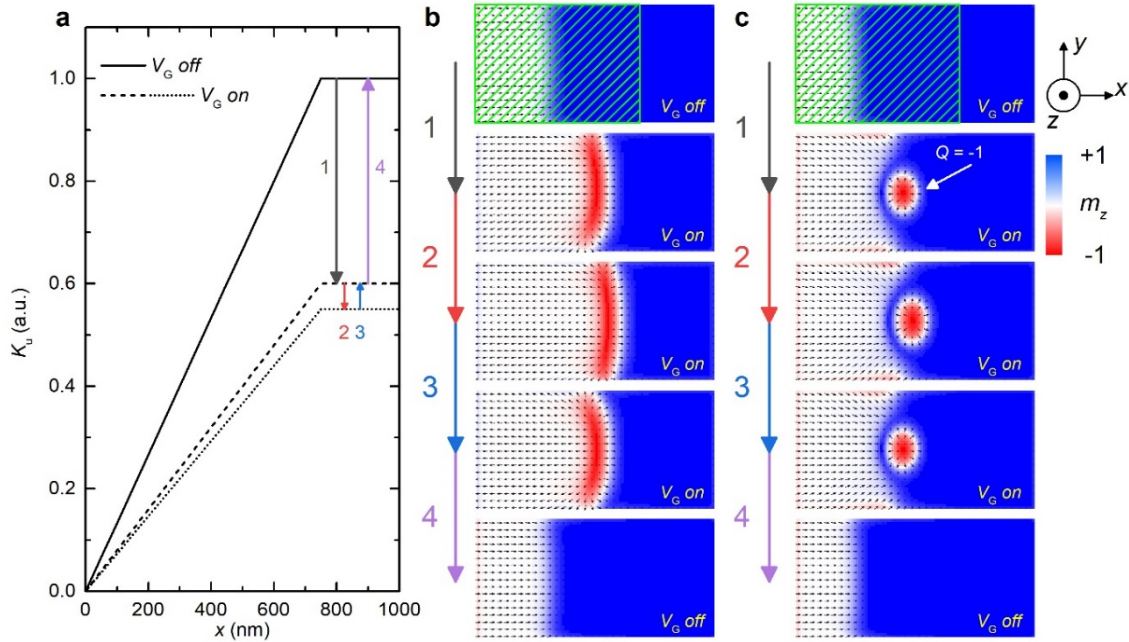

**Figure 6. Simulation of the EF-induced creation and motion of a chiral domain wall. a**, The $V_G$-dependent profile of the PMA ($K_u$). The PMA linearly increases from $K_u = 0$ at $x = 0$ nm to $K_u = 1.0K$ at $x = 750$ nm when $V_G$ is off. When $V_G$ is on, the PMA linearly increases from $K_u = 0$ to $K_u = 0.6K$. When $V_G$ is further increased, the PMA linearly increases from $K_u = 0$ to $K_u = 0.55K$. **b**, The EF-induced creation and motion of a chiral domain wall toward the right and left at $H_z = 0$ mT. The numbers in **b** and **c** indicate the evolution of the chiral spin textures when the PMA is changed by the EF. The snapshots show the relaxed states corresponding to each change of the PMA. **c**, The EF-induced motion of a chiral bubble toward the right and left at $H_z = 0$ mT. A relaxed chiral bubble with $Q = -1$ is given as the initial state when $V_G$ is off. The color scale represents the out-of-plane magnetization component $m_z$. The in-plane magnetization components ($m_x$, $m_y$) are denoted by arrows.